\documentclass[10pt,twocolumn,a4] {article}
\usepackage{polski}
\usepackage[english]{babel}
\usepackage{graphicx,colordvi}
\usepackage{amssymb}
\usepackage{xcolor}
\usepackage{alltt}
\title{\bf Potential energy surfaces and fission fragment mass yields of even-even superheavy nuclei\footnote{Supported by the Polish National Science Center (Grant No. 2018/30/Q/ST2/00185) and by the National Natural Science Foundation of China (Grant No. 11961131010 and 11790325).}}

\author{Pavel V. Kostryukov$^1$, Artur Dobrowolski$^1$, 
Bo\.zena Nerlo-Pomorska$^1$, Micha{\l} Warda$^1$,\\ 
Zhigang Xiao$^2$\footnote{Email: xiaozg@mail.tsinghua.edu.cn}, 
Yongjing Chen$^3$, Lile Liu$^3$, Jun-Long Tian$^4$, Krzysztof Pomorski$^1$\footnote{Email Krzysztof.Pomorski@umcs.pl}\\
{\footnotesize $^1$ Institute of Physics, Maria Curie Sk{\l}odowska University, 20-031 Lublin, Poland}\\
{\footnotesize $^2$ Department of Physics, Tsinghua University, Beijing 100084, China}\\
{\footnotesize $^3$ Institute of Atomic Energy, Beijing 102413, China}\\
{\footnotesize $^4$ School of Physics and Electrical Engineering, Anyang Normal University, Anyang 455000, China}}

\date{\today}

\begin{document}
\onecolumn

\maketitle

\begin{center} \begin{minipage}{16cm} {\bf Abstract:}  

Potential energy surfaces and fission barriers of superheavy nuclei are analyzed in the macroscopic-microscopic model. The Lublin-Strasbourg Drop (LSD) is used to obtain the macroscopic part of the energy, whereas the shell and pairing energy corrections are evaluated using the Yukawa-folded potential. A standard flooding technique has been used to determine the barrier heights. It was shown the Fourier shape parametrization containing only three deformation parameters reproduces well the nuclear shapes of nuclei on their way to fission. In addition, the non-axial degree of freedom is taken into account to describe better the form of nuclei around the ground state and in the saddles region. Apart from the symmetric fission valley, a new very asymmetric fission mode is predicted in most superheavy nuclei. The fission fragment mass distributions of considered nuclei are obtained by solving the 3D Langevin equations.



\end{minipage}
\end{center}

\twocolumn
\normalsize

\section{Introduction}

The theoretical study of the properties of superheavy nuclei (SHN) 

are of great importance since the confrontation with the experimental data offers a stringent test of any nuclear model. Most nuclear theories can achieve a fair description of masses and other properties of nuclei close to the $\beta$-stability line. However, their predictions often deviate when moving to nuclei far from stability, e.g., the SHN region. Apart from the apparent interest to better understand the involved physics, the predictive power of these theoretical approaches plays an essential role in guiding the challenging experimental quest for the so-called SHN island of stability and limits of existence of bound nuclei.

The superheavy nuclei \cite{HHO13,Khy13} with charge number Z up to 118, have been produced by two kinds of fusion reactions: cold fusion at GSI Germany \cite{HMu00,Hof11} and RIKEN Japan \cite{Mor07} using doubly magic target $^{208}$Pb or its neighbour $^{209}$Bi, and hot fusion with the $^{48}$Ca projectile at JINR Dubna Russia and Livermore Nat. Lab. USA \cite{Oga10,Oga15}. Further attempts to synthesize Z= 119,120 isotopes are in progress (confer e.g. \cite{Hof16,Hof17}). Several theoretical studies were also performed in the last years studying properties, possible decay modes, and fission fragment mass yields \cite{CIO15,CIO19,IZU20,JKS20,MLZ20,MJV19,CZL18}, as well as synthesis chances \cite{LIK20,HAA21,SCK19,CKW21} of nuclei in this region.

The present paper is a continuation of our work on fission fragment mass yields of even-even Ra-Th and actinide nuclei \cite{PDH20,PBK21,LCW21}. We have shown there the macroscopic-microscopic (mac-mic) model based on the LSD formula \cite{PDu03} and the Yukawa-folded single-particle potential \cite{DNi76}  describes well both fission barrier heights and the fission fragment mass yields (FMY). In the present investigation, we aim to predict, using the same set of the model parameters, the fission barrier heights and the FMY's of the SHN. The calculation is made in the 4D Fourier deformation parameter space \cite{PNB15,SPN17}.

The Born-Oppenheimer+Wigner model used in Ref.~\cite{PBK21} is not foreseen to describe a situation in which one deals with two competing fission modes. So, in the present research, we evaluate the FMY by solving the 3D Langevin equations like it was done in Ref.~\cite{LCW21} for some actinides. This change is since, in some isotopes of SHN, our mac-mic model predicts two well-separated fission valleys: one corresponding to the symmetric fission and the other asymmetric one, which leads to the heavy fragments around $^{208}$Pb.

The paper is organized in the following way. After the introduction, in Section 2, we present details of the theoretical models used in the present study. Then, in Section 3, we show the collective potential energy surface evaluated within the mac-mic model for the selected isotopes and our estimates of the fission barrier heights. The estimated FMY's are presented in Section 4. Conclusions and perspectives of further investigations are at the end of the article.

\section{Theoretical model}
\label{theory}

Potential energy surfaces (PES) of fissioning nuclei are studied within the mac-mic model in the four-dimensional space spanned by deformation parameters describing the elongation, left-right asymmetry, neck, and non-axiality of the nucleus. Detailed study of the evaluated PES for the SHN allows estimating the equilibrium deformations, possible shape coexistence and shape isomers, fission barrier height, and fission valleys. The dissipative fission dynamics at the obtained PES are described here by the Langevin equations set, which estimates the possible fission modes and corresponding FMY. Below we briefly present the main ingredients of our model.

\subsection{Fourier nuclear shape parametrization}
\label{fs}

The axial symmetric shape-profile function of a fissioning nucleus written in cylindrical coordinates $(\rho,z)$ is expanded in a Fourier series \cite{PNB15,SPN17} as:
\begin{equation}
\begin{array}{ll}
\frac{\rho^2_s(u)}{R_0^2} =& a_2\cos(u)+a_3\sin(2u)+a_4\cos(3u)\\[1ex]
    &+a_5\sin(4u) + a_6\cos(5u)+\dots~,
\end{array}
\label{Fourier}
\end{equation}
where $R_0$ is the radius of spherical nucleus and $u=\pi/2\cdot(z-z_{\rm sh})/z_0$ with $-z_0+z_{sh}\le z\le z_0+z_{\rm sh}$. The volume conservation condition gives $z_0=R_0\pi/(a_2-a_4/3+a_6/5-\dots)/3$. The shift of $z$-coordinate $z_{\rm sh}$ ensures that the centre of mass is located at the origin of the coordinate system. One can use the Fourier expansion coefficients as free deformation parameters, but it is more effective to make their combinations $\{q_n\}$, called {\it optimal coordinates} \cite{SPN17}:
\begin{equation}
\left\{
\begin{array}{ll}
 q_2=a_2^{(0)}/a_2 - a_2/a_2^{(0)}~,& q_3=a_3~, \\[1ex]
 q_4=a_4+\sqrt{(q_2/9)^2+(a_4^{(0)})^2}~,&\\[1ex]
 q_5=a_5-(q_2-2)a_3/10~,&  \\[1ex]
 q_6=a_6-\sqrt{(q_2/100)^2+(a_6^{(0)})^2} &\;\;.
\end{array} \right.
\label{qi}
\end{equation}
The deformation parmeters $q_n(\{a_i\})$ were chosen in such a way that the liquid-drop energy as a function of the elongation $q_2$ becomes minimal along a trajectory that defines the liquid-drop path to fission. The $a^{(0)}_{2n}$ in Eq.(\ref{qi}) are the expansion coefficients of a spherical shape given by $a^{(0)}_{2n}=(-1)^{n-1}32/\pi^3/(2n-1)^3$.

Non-axial shapes can easily be obtained assuming that, for a given value of the $z$-coordinate, the surface cross-section has the form of an ellipse with half-axes $a(z)$ and $b(z)$ \cite{SPN17}:
\begin{equation}
   \hspace{-0.3cm}  \varrho^2(z,\varphi) = \rho^2_s(z) 
                \frac{1-\eta^2}{1+\eta^2+2\eta\cos(2\varphi)}
   \hspace{0.3cm} \mbox{with} \hspace{0.3cm} 
 \eta = \frac{b-a}{a+b}~,
\label{eta}
\end{equation}
where the parameter $\eta$ describes the non-axial deformation of the nuclear shapes. The volume conservation condition requires that $\rho_s^2(z)=a(z)b(z)$.

\subsection{Macroscopic-microscopic model}
\label{mm}

In the mac-mic method, proposed first by Myers and \'Swi{\k a}tecki \cite{MSw66}, the total energy of the deformed nucleus is equal to the sum of a macroscopic (liquid-drop type) energy and the quantum energy correction for protons and neutrons generated by shell and pairing effects
\begin{equation}
 E_{\rm tot} = E^{}_{\rm LSD}+ {\rm E_{shell}} + {\rm E_{\rm pair}} \; .
\label{eq01}
\end{equation}
The LSD model \cite{PDu03} which reproduces well all experimental masses and fission barrier heights, is used in this study to evaluate the macroscopic part of the energy. The shell corrections are obtained by subtracting the average energy $\widetilde E$ from the sum of the single-particle (s.p.) energies of occupied orbitals
\begin{equation}
 {\rm E_{\rm shell}} = \sum_k e_k - \widetilde E \; .
\label{eq02}
\end{equation}
As the s.p. energies $e_k$, we have taken the eigenvalues of a mean-field Hamiltonian with the Yukawa-folded s.p. potential \cite{DNi76}. The average energy $\widetilde E$ is evaluated using the Strutinsky prescription \cite{Str66, NTS69} with a 6$^{\rm th}$ order correction polynomial. The pairing energy correction is determined as the difference between the BCS energy \cite{BCS57}, and the s.p. energy sum from which the average pairing energy \cite{NTS69} is subtracted
\begin{equation}
 E_{\rm pair} = E_{\rm BCS} - \sum_{k} e_k
            - \widetilde{E}_{\rm pair}\;.
\label{Epair}
\end{equation}
In the BCS approximation, the ground-state energy of a system with an even number of particles is given by
\begin{equation}
E_{\rm BCS} = \sum_{k>0} 2e_k v_k^2 - G(\sum_{k>0}u_kv_k)^2 - G\sum_{k>0} v_k^4
 -{\cal E}_0^\phi\;,
\label{EBCS}
\end{equation}
where the sums run over the pairs of s.p. levels belonging to the pairing window defined below. The coefficients $v_k$ and $u_k=\sqrt{1-v_k^2}$ are the BCS occupation amplitudes, and ${\cal E}_0^\phi$ is the energy correction due to the particle number projection done in the GCM+GOA approximation \cite{GPo86}
\begin{equation}
{\cal E}_0^\phi=\frac{\sum\limits_{k>0}[ (e_k-\lambda)(u_k^2-v_k^2)
        +2\Delta u_k v_k +Gv_k^4] / E_k^2}{\sum\limits_{k>0} E_k^{-2}}\;.
\label{Ephi}
\end{equation}
Here $E_k=\sqrt{(e_k-\lambda)^2+\Delta^2}$ are the quasi-particle energies and $\Delta$ and $\lambda$ the pairing gap and the Fermi energy, respectively. The average projected pairing energy, for a pairing window, symmetric in energy with respect to the Fermi energy, of width $2\Omega$, is equal to
\begin{equation}
\begin{array}{l}
 \widetilde{E}_{\rm pair}=\displaystyle{-\frac{1}{2}\,\tilde{g}\,
 \tilde{\Delta}^2+\frac{1}{2}\tilde{g}\,G\tilde{\Delta}\,
 {\rm arctan}\left(\frac{\Omega}{\tilde\Delta}\right)
  -\log\left(\frac{\Omega}{\tilde\Delta}\right)\tilde{\Delta}}\\[3ex]
~~~~~~~~~\displaystyle{
+\frac{3}{4}G\frac{\Omega/\tilde{\Delta}}{1+(\Omega/\tilde{\Delta})^2}/
  {\rm arctan}\left(\frac{\Omega}{\tilde{\Delta}}\right)-\frac{1}{4}G }~ ,
\end{array}
\label{Epavr}
\end{equation}
where $\tilde{g}$ is the average single-particle level density and $\tilde\Delta$ the average paring gap corresponding to a pairing strength $G$ 
\begin{equation}
\tilde\Delta=\displaystyle{2\Omega\exp\left(-\frac{1}{G\tilde{g}}\right)}~.
\label{Davr}
\end{equation}
The pairing window for protons or neutrons contains $2\sqrt{15\cal N}$ ($\cal N =\,$N or Z) s.p. levels closest to the Fermi energy states. For such a window, the paring strength approximated in Ref.~\cite{PPS89} is given by the following expression:
\begin{equation} 
 G = \displaystyle{\frac{g_0}{{\cal N}^{2/3} \, A^{1/3}}}~.
\label{Gpair}
\end{equation}
The same value $g_0=g_0^p=g_0^n=0.28 \hbar \omega_0$ is taken for protons and neutrons, where $\hbar \omega_0 = 41$\,MeV$/A^{1/3}$ is the nuclear harmonic oscillator constant.

In our calculation, the single-particle spectra are obtained by diagonalization of the s.p. Hamiltonian with the Yukawa-folded potential \cite{DNi76, DPB16} with the same parameters as used in Ref.~\cite{MNi95}.


\subsection{Multidimensional Langevin equation}
\label{dd}

To study the fission dynamics of atomic nuclei, we use the formalism of the Langevin equations, which determine the motion of the nucleus in the multidimensional space of deformation parameters $q_i$ Eq. (\ref{qi}). Such system of coupled equations is similar to the canonical Hamilton equations with friction but  it contains in addition a stochastic force. The Langevin equations can be written as follows (confer e.g. Ref.~\cite{KPo12}):
\begin{equation}
\left\{
\begin{array}{ll}
\displaystyle{\frac{dq_i}{dt}=}& \displaystyle{\sum\limits_{j} \left[\mathcal{M}^{-1}\right]_{ij}p_j,}\\[3ex]
\displaystyle{\frac{dp_i}{dt}=}&\displaystyle{-\frac{\partial V}{\partial q_i}-\frac{1}{2}\sum\limits_{jk} {\left[\frac{\mathcal{M}^{-1}}{\partial q_i}\right]_{jk}}p_jp_k} \\[3ex]
&\displaystyle{+ \sum\limits_{jk}\gamma_{ij} \left[\mathcal{M}^{- 1} \right]_{jk}p_k + \sum_{j} g_{ij} \Gamma_j ~ ,}
\end{array} 
\right. ~
\label{Leq}
\end{equation}
where $p_i$ is the conjugated momentum corresponding to the coordinate $q_i$, while $\mathcal{M}_{ij}(q)$ and $\gamma_{ij}(q)$ are the inertia and the friction tensors, respectively, and $V(q)$ is the potential energy of the fissioning nucleus. 

The inertia tensor is calculated within the incompressible and irrotational liquid drop model using the Werner-Wheeler approximation \cite{DSN76}. For the nuclear surface described by the function $\rho^2_s(z,q)$ (Eq.~(\ref{Fourier})) the inertia tensor is given by the following formula \cite{KPo12}:
\begin{equation}
 \mathcal{M}_{ij}(q) = \pi \rho_m \int\limits_{z_{min}}^{z_{max}} \rho^2_s(z,q)
  \left[ A_i A_j + \frac{1}{8}\rho^2_s(z, q )A_i' A_j' \right] dz ~.
\label{Iten}
\end{equation}
Here $\rho_m = M_0/(\frac{4}{3}\pi R_0^3)$ is the density of nucleus. The velocity expansion coefficients $A_i$ in Eq.~(\ref{Iten}) have the following form:
\begin{equation}
 A_i=\frac{1}{\rho^2_s(z, q)}\frac{\partial}{\partial q_i}\int_{z}^{z_{max}}
     \rho^2_s (z', q) dz' ~,
\label{Avec}
\end{equation}
and $A_i'=\partial A_i/\partial z$. 

During the fission process the temperature of nucleus changes due to the existence of the friction forces. To take this effect into account one has to use another type of the potential, known as the Free Helmholtz energy $F(q)$, depending on the temperature, instead the temperature independent potential $V(q)$ in the Langevin equation (\ref{Leq}) (confer e.g. \cite{KPo12}):
\begin{equation}
	 F(q) = V(q) - a(q)T^2~,
\label{Efree}
\end{equation}  
where $T$ is the temperature of nucleus
\begin{equation}
 T=\sqrt{E^*/a(q)}~.
\label{Temp}
\end{equation} 
Here $E^*$ is the thermal (statistical) excitation energy of nucleus and $a(q)$ is the density of s.p. levels. In our calculation the parameter $a(q)$ is taken from Ref.~\cite{NPB06}.

The collective potential Eq.~(\ref{eq01}) in the mac-mic approximation is given by the sum of the macroscopic and the microscopic parts $V=V_{\rm mac}+V_{\rm mic}$. The first term is almost temperature independent at low excitation energies, while the  temperature dependence of the microscopic energy correction may be approximated as follows \cite{NPB06}:
\begin{equation}
	 V_{\rm mic}(q, T) = \displaystyle{\frac{V_{\rm mic}(q, T = 0)}{1 + e^{(1.5 - T)/0.3}}}~.
\label{Vmic}
\end{equation}

Also the friction forces vary with temperature: they vanish in a cold system and grow with the excitation of nucleus. We take into account their temperature dependence using the following function: 
\begin{equation}
\gamma^{\rm mic}_{ij}=\displaystyle{\frac{0.7\cdot\gamma^{\rm wall}_{ij}}
                      {1+e^{(0.7-T)/0.25}}}~,
\label{fricT}
\end{equation}
which approximates the estimates done in Ref.~\cite{IPo96}. Here the friction tensor $\gamma^{\rm wall}_{ij}$ is given by the wall-formula \cite{BBN78}:
\begin{equation}
	\gamma^{\rm wall}_{ij} = \frac{\pi}{2} \rho_m \bar{v} \int\limits_{z_{min}}^{z_{max}} \frac{\partial \rho^2_s}{\partial q_i} \frac {\partial \rho^2_s}{\partial q_j} \left[ \rho^2_s + \frac{1}{4} \left (\frac {\partial \rho^2_s}{\partial z} \right)^2 \right]^{-1/2} dz,
\label{Ften}
\end{equation}
where $\bar{v}$ is the average internal velocity of nucleons in the nucleus, and its value is related to the Fermi velocity $v_F$ as $\bar{v} = \frac{3}{4} v_F$. 

The last term in the second equation in Eq. (\ref{Leq}) represents the random Langevin force which amplitude $g_{ij}$ being the square-root of the diffusion tensor $D_{ij}$ and $\Gamma= \xi \cdot \sqrt{\tau})$ is the time-dependent random function,  where $\xi$ is defined as a random Gaussian distribution with properties similar to those of white noise:
\begin{equation}
     \bar{\xi} = 0, \: \bar{\xi}^2 = 2~,
\label{Rforce}
\end{equation}
and $\tau$ is time steps used when solving the Langevin equations.

The diffusion tensor is obtained using the Einstein relation
\begin{equation}
    D_{ij}=\sum_k g_{ik} g_{jk} = \gamma_{ij} \cdot T~,
\label{Erel}
\end{equation}
Unfortunately the Einstein relation which is fulfilled in systems having relatively high temperature does not take into account the quantum fluctuations present at low excitation of nuclei. In order to extend the application of the Langevin equations to low energy fission one replaces the temperature $T$ in Eq.~(\ref{Erel}) by the effective temperature $T^*$ introduced in ~\cite{WWH81,HKi98}:
\begin{equation}
     T^* = \frac{E_0}{2} \coth \frac{E_0}{2 T}~,
\label{Tstar}
\end{equation}   
where $E_0$ corresponds to the zero-point energy of collective vibrations which is of the order 1 MeV. 
\begin{figure}[h!]
\includegraphics[width=\columnwidth]{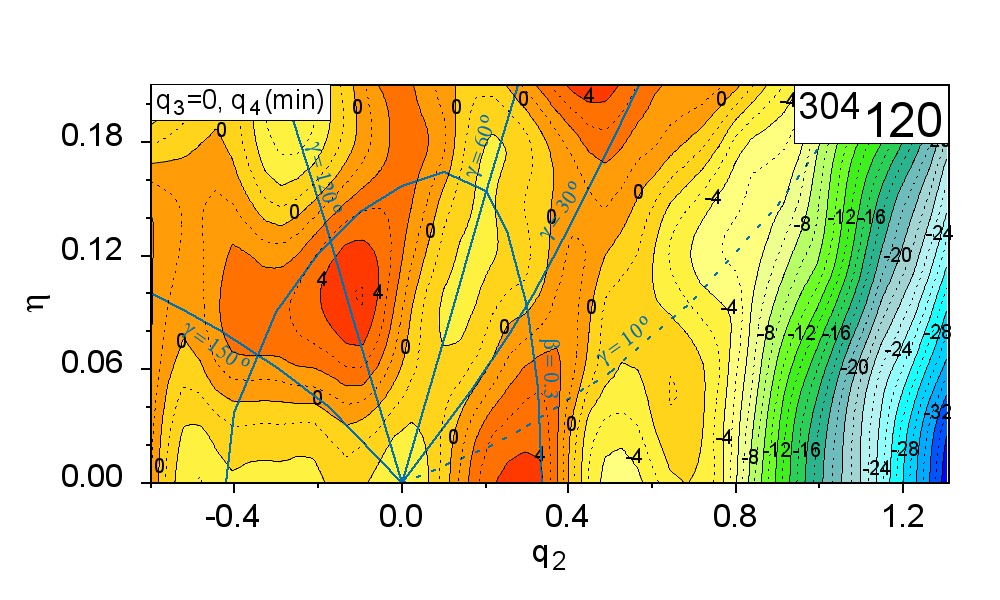}
\includegraphics[width=\columnwidth]{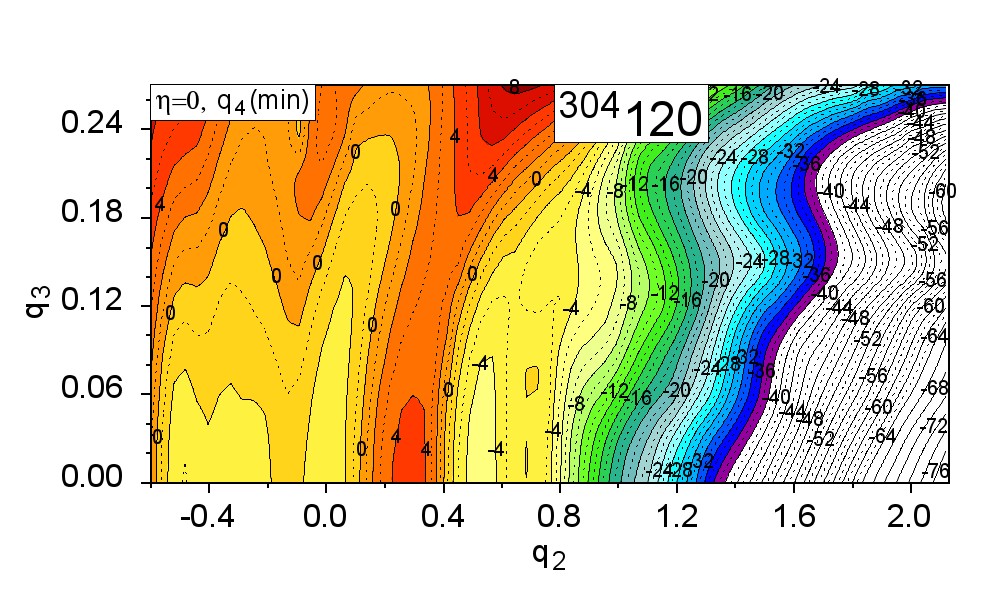}
\caption{Potential energy surface of $^{304}$120 isotope minimized with respect to $q_4$ at the $(q_2,\eta)$ (top) and $(q_2,q_3)$ (bottom).
The lines corresponding to $\gamma\,=\,10^{\rm o},\,30^{\rm o},\,60^{\rm o},\,120^{\rm o}$, and $150^{\rm o}$ as well as the line corresponding to $\beta$=0.3 are marked in the $(q_2,\eta)$ maps.}
\label{So304}
\end{figure}

The irrotational flow inertia tensor Eq.~(\ref{Iten}) and the wall friction tensor Eq.~(\ref{Ften}) are evaluated using the Fortran codes published in Ref.~\cite{BNP19}.


\section{Potential energy surfaces}
\label{PES}

The nuclear potential energies of even-even superheavy nuclei are evaluated in the equidistant grid in the 4D collective space built on the $q_2, ~q_3, ~q_4$, and $\eta$ deformation parameters. The total energy function of a nucleus is obtained as described in Sec.~\ref{mm}. It is rather challenging to present graphically such a 4D object. So, in the following, we show only some selected 2D cross-sections of the full PES. 

In Appendix, the cross-sections $(q_2,\eta)$ (l.h.s. column) and $(q_2,q_3)$ (r.h.s. column) of the 4D potential energy surface of elements Ds to Z=120 are shown. For each element, five pairs of maps corresponding to different isotopes are displayed. To better understand the properties of the cross-sections presented in the Appendix, we show below as an example two cross-sections for $^{304}$120 isotope.

In the top panel of Fig.~\ref{So304} the ($q_2,\,\eta$) cross-section of the PES of $^{304}$120 nucleus is presented. Each energy point of the plot is minimized with respect to the neck parameter $q_4$. Solid green lines are drawn in the figure to guide the eye. The lines correspond to approximate positions of the frequently used $\beta$ and $\gamma$ Bohr deformation parameters describing a spheroid \cite{Boh52}. We write here 'approximate' as the shapes considered in this paper are much richer than the spheroidal one. The lines $\gamma=30^{\rm o}$ and $150^{\rm o}$ correspond to the largest non-axial deformation, while those for $\gamma=0,~120^{\rm o}$ to the prolate shapes of the nucleus and $\gamma=60^{\rm o}$ and $180^{\rm o}$ to the oblate forms. In addition, the green dashed line shows the $\gamma=10^{\rm o}$ direction. To give the reader an orientation about the elongation of the nucleus, the line corresponding to $\beta=0.3$ is displayed. The bottom panel shows the cross-section $(q_2,\,q_3)$ of the PES of $^{304}$120 for the axial symmetric case ($\eta=0$). Each energy point of the map is minimized to $q_4$ deformation. The layers in both maps correspond to the total energy of the nucleus measured relative to the LSD energy spherical nucleus. The distance between the solid line layers is 2 MeV, while the dotted lines correspond to the half-layers.

As one can see, the nucleus $^{304}$120 is spherical in the ground-state and possesses two oblate shape isomers of comparable energy. One can even speak here on the shape coexistence. The least energy path to fission reminds the situation which one observes in rotating nuclei \cite{MKK01}. First, it becomes oblate and then via triaxial shapes goes to fission. The effect of the non-axial degree of freedom ends around $q_2=0.8$ and $\eta=0.12$ (or $\gamma=10^{\rm o}$). At larger deformations, the $\eta$ degree of freedom does not play an essential role. Therefore, it can be omitted when one discusses fission dynamics. The fission barrier reduction due to the breaking of the axial symmetry in $^{304}$120 is around 3 MeV. 

In the bottom panel of Fig.~\ref{So304} the $(q_2,\,q_3)$ cross-section of the PES of $^{304}$120 is shown. One can see there that taking into account the left-right (octupole like) asymmetry diminishes the fission barrier of $^{304}$120 by approximately 1 MeV. At smaller elongation $q_2<0.3$ the minimal energy corresponds to $q_3=0$. Beyond the saddle point at $(q_2\approx 0.4,\,q_3\approx 0.13,\,\eta\approx 0.14)$ the role of the octupole deformation becomes more and more important. Two fission valleys are formed with the growing elongation of the nucleus. One is left-right symmetric ($q_3=0)$ which goes through a local minimum at $q_2\approx 0.5$ and ends at scission at $q_2\approx 2.3$ giving the symmetric in mass fission fragment distribution. The second valley for $q_3\approx 0.2$ leads the very asymmetric fission with the masses of the heavy fragments concentrated around A=208. It is the effect of the double magic $^{208}$Pb. The possibility of the existence of such decay mode of the SHN was also foreseen by Poenaru et al. \cite{PSG18}, and by Warda with co-workers \cite{WZR18} as a cluster emission. In Ref.~\cite{WZR18} it was shown in addition that such mode may be treated as a super-asymmetric fission.

The maps presented in the Appendix for $^{254-262}$Rf, $^{258-266}$Sg, $^{264-272}$Hs, $^{276-284}$Ds, $^{278-286}$Cn, $^{282-290}$Fl, $^{286-294}$Lv, $^{290-298}$Og, and $^{294-302}$120 illustrate well the interplay between the non-axial and left-right asymmetric deformations of fissioning super-heavy nuclei. It is only a selection of similar results which we have obtained for 18 even-even isotopes chains of each element with 104$\leq$Z$\leq$126. Such a broad range of superheavy isotopes for which we have made the calculation is not only interesting from the point of view of their possible synthesis, but also it is important for astrophysical models (confer, e.g. Ref.~\cite{SAG19, SAG20, LGB21}).
\begin{figure}[h!]
\includegraphics[width=0.95\columnwidth]{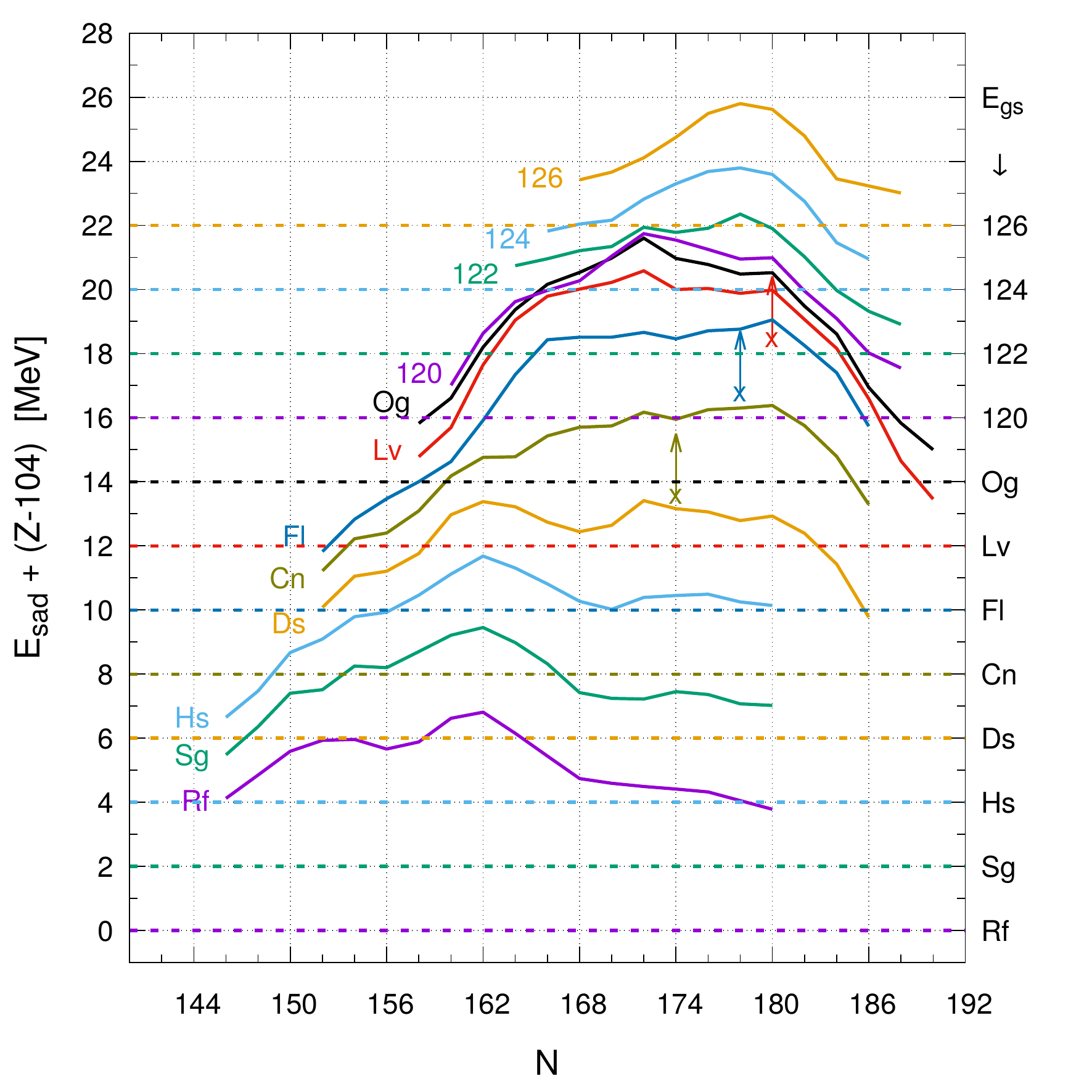}
\caption{Fission barrier heights of even-even superheavy nuclei evaluated in our 4D mac-mic model. The ground-state energy for Rf isotopes is taken as zero, while for each subsequent element, it is shifted by 2 MeV. The barrier plots are also shifted by the same amount. The ground state energy values corresponding to each element are marked at the r.h.s. vertical scale. The experimental data for the lower limit of the barrier heights \cite{IOZ02} are marked by crosses and arrows of the same colors as element symbols.}
\label{barr}
\end{figure}

Of course there is not sufficient place in a regular article to discuss the details of $(q_2,\,q_3 )$ and $(q_2,\,\eta)$ maps for each isotope. So, we present in Fig.~\ref{barr} the fission barrier height systematics only. 

The barrier heights are evaluated using the flooding technique in the 4D space. The results for each element starting from Rf to Z=120 are drawn with different colors. The ground-state energy for Rf isotopes is taken equal to zero. At the same time, for each subsequent element, it is shifted by 2 MeV, i.e. one has to subtract $(Z\,-\,104)$ MeV from the displayed values of the barrier height of element Z. Or, in another way, one has to observe the difference between the ground state energy (drawn with the dashed line of the same color) of each element and an appropriate solid curve.   

The maximal barrier heights for each element varies from around 7 MeV for Rf to Ds nuclei, reaching  9 MeV for $^{294}$Fl, and then decrease up to element Z=120 having its upper value around 5 MeV only. Then the effect of the semi-magic proton number appears, and the barrier height grows with Z number reaching a value 3.76 MeV for the hypothetical isotope $^{304}$126. 
\begin{figure}[h!]
\includegraphics[width=\columnwidth]{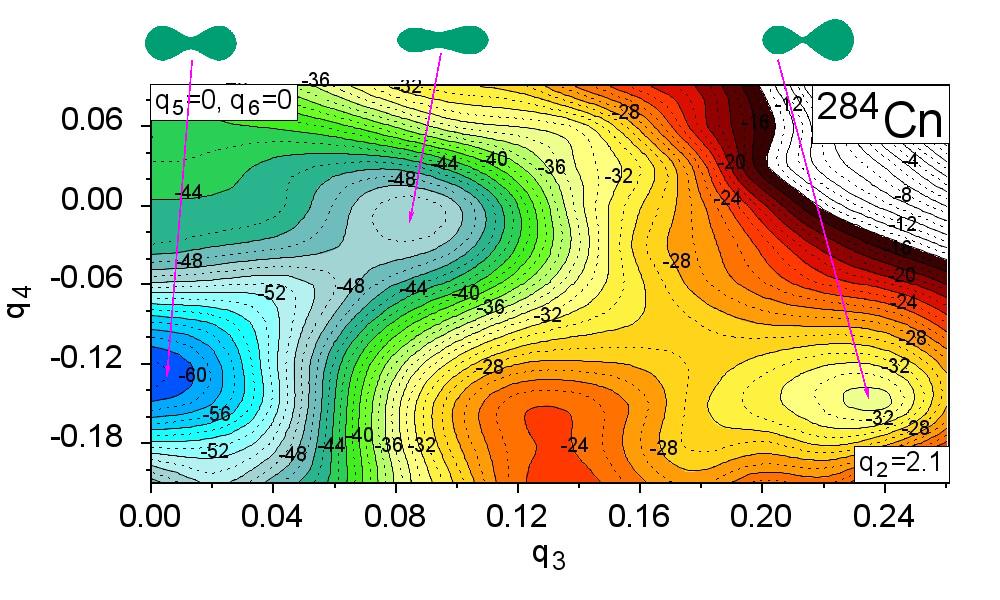}\\
\includegraphics[width=\columnwidth]{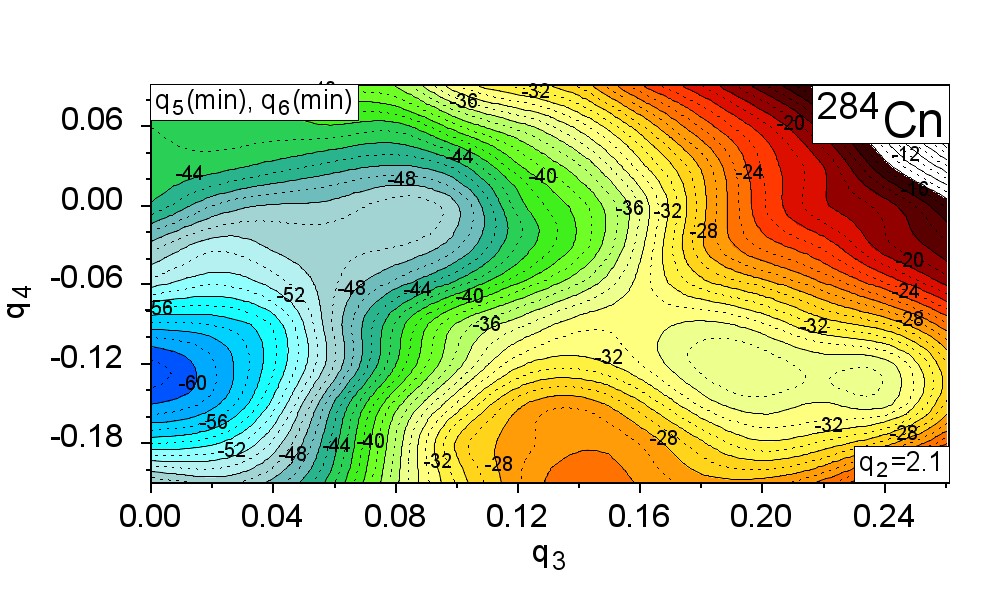}
\caption{The $(q_3,q_4)$ cross-section of the PES of $^{284}$Cn taken at the elongation $q_2=2.1$. The upper plot corresponds to the case when the higher order deformations are not taken into account while the bottom one shows the PES minimized with respect to $q_5$ and $q_6$.}
\label{e34cross}
\end{figure}
\begin{figure}[h!]
\includegraphics[width=\columnwidth]{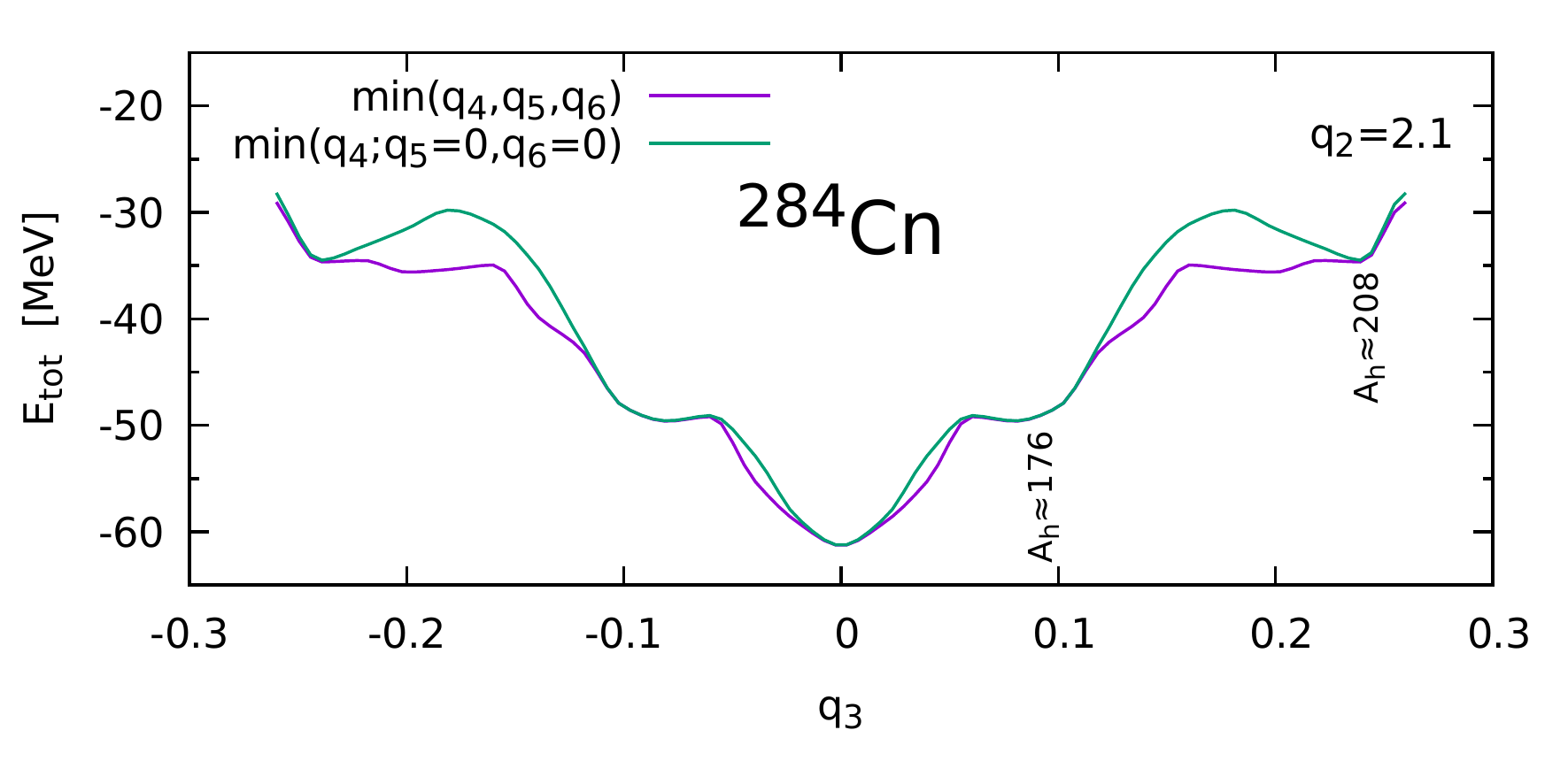}
\caption{PES cross-section corresponding to the elongation $q_2=2.1$ as a function of the $q_3$ deformation. The thick violet line corresponds the PES minimized with respect to $q_4$, $q_5$, and $q_6$, while the thin green line to the case when $q_5$ and $q_6$ are not taken into account.}
\label{cross} 
\end{figure}

The question arises if our 4D deformation space ($q_2,\,q_3,\,q_4,\,\eta$) is sufficient to describe the fission barriers and valleys as some other authors take much more deformation parameters into account to obtain similar results (confer, e.g. Ref.~\cite{JKS20}). It was shown in Refs.~\cite{SPN17,PNB20} that the 4D Fourier shape parametrization used in the present research describes very well the main features of the potential energy of fissioning nuclei. The influence of the high orders Fourier expansion terms on estimates of the PES of fissioning actinide nuclei was studied in Ref.~\cite{PNB20}. It was found there that this effect is rather small. In the superheavy region of nuclei, the change of the PES due to the $q_5$ and $q_6$ deformations is either not large, even at elongations of the nucleus close to the scission configuration. It can be seen in Fig.~\ref{e34cross}, where the ($q_3,\,q_4$) cross-section of the PES of $^{284}$Cn at the quite significant  elongation $q_2=2.1$ is shown. The upper panel corresponds to the case when the deformations $q_5$ and $q_6$ are not taken into account, while the bottom one shows the PES minimized concerning these high order deformations. The shapes of the nucleus in the local minima are also shown in the upper panel. Comparing both maps in the vicinity of the minimum at $q_3=0$ and $q_4=-0.13$ corresponding to the symmetric fission valley, one can see that the PES is almost unchanged. Some effect of $q_5$ and $q_6$ is seen around the local minimum at $q_3=0.23$ and $q_4=-0.13$ corresponding to the very asymmetric fission valley. Taking into account the $q_5$ and $q_6$ deformations makes this minimum broader and reduces the barrier height separating both minima. 

The PES cross-section of $^{284}$Cn corresponding to the minimal energy as function of $q_3$ for the constant elongation $q_2=2.1$ is drawn in Fig.~\ref{cross}. The thick violet line corresponds to the case when each point of the curve is minimized with respect to $q_4$, $q_5$, and $q_6$, while the thin green line shows the potential energy when the effect $q_5$  and $q_6$ is neglected. One can see the influence of the higher-order deformations on the energy values in the minima is hardly observed. 

Our group presented similar estimates of the SHN barrier height in Ref.~\cite{PNB18}. The main difference between those and present results origins mainly from a better and more accurate description of the pairing correlations effect (see Eqs. (\ref{EBCS}-\ref{Epavr})), as well as using denser and more extended 4D mesh in the deformation parameter space. 
\begin{figure}[h!]
\centerline{\includegraphics[width=0.7\columnwidth]{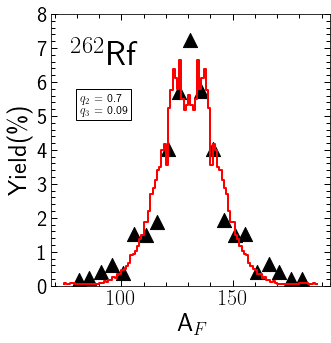}}
\caption{Fission fragment mass yield (red solid line) estimate for $^{262}$Rf nucleus. The experimental (for the spontaneous fission case) data (black triangles) are taken from Ref.~\cite{Lan96}.}
\label{Rf262}
\end{figure}
\begin{figure}[h!]
\centerline{\includegraphics[width=0.8\columnwidth]{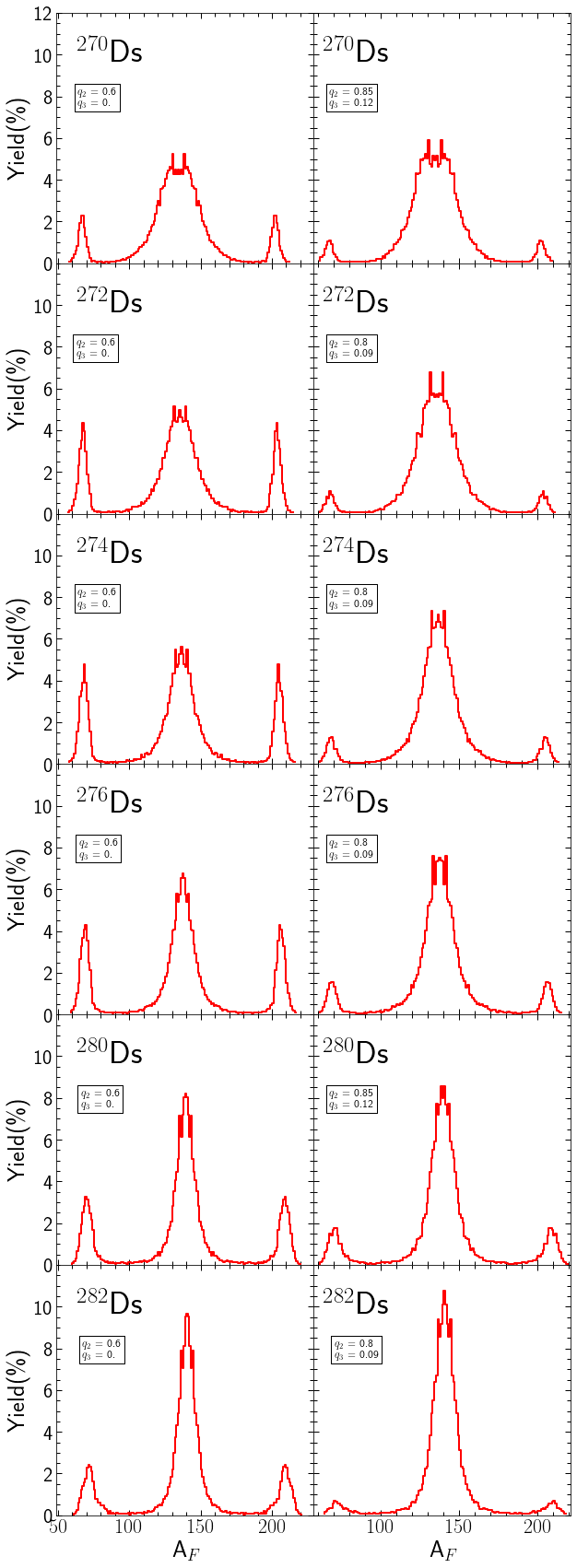}}
\caption{Fission fragment mass yields of six Ds isotopes. The l.h.s. column
corresponds to the case when the Langevin trajectories begin in the vicinity
of the saddle point (low energy fission) while the r.h.s. columns present the estimates made for the spontaneous fission when the trajectories
begin around the exit point from the barrier.}
\label{Dsy}
\end{figure}

 
\section{Fission fragment mass yields}

Now, having introduced the details of generating the PES's, we may switch to the statistical approach based on the Langevin formalism to find the fragment mass distributions of the fissioning nucleus. The set of coupled Langevin equations defined in the Fourier deformation space, leading to the bundle of stochastic trajectories between the ground state and a scission configuration on the scission surface, are already described in Subsection 2.3.

All deformation dependent transport coefficients in Eq.~(\ref{Leq}) were stored for each nucleus in equidistant ($\Delta q_2=0.05,\Delta q_3=0.03,\,\Delta q_4=0.03$) mesh points in the 3D Fourier deformation parameters space. The values of the PES and the transport function and their derivatives between grid points are obtained using the Gauss-Hermitian approximation method~\cite{Pom06}. The non-axial degree of freedom $\eta$ in Eq.~(\ref{eta}) is not taken into account as its role at a large elongation of nuclei is negligible.
\begin{figure}[htb]
\centerline{\includegraphics[width=0.8\columnwidth]{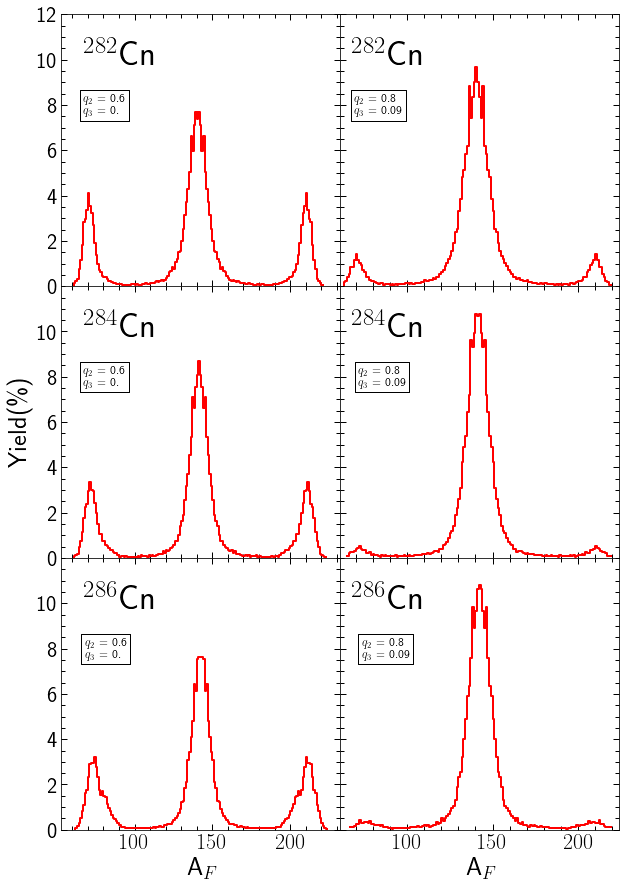}}
\caption{The same as in Fig.~\ref{Dsy} but for Cn isotopes.}
\label{Cny}
\end{figure}

The Langevin calculation of each $n^{\rm th}$ trajectory begins at a starting point $\{q^{\rm start}_n\}$. Since the starting point can be chosen, to some extent, arbitrarily, a natural question arises, which configuration one has to take as the beginning of the trajectories?  Location of the ground state? First or second saddle point? Or the exit-point after tunnelling of the potential energy barrier in the spontaneous fission case? To answer these questions, we discuss the following two types of starting points: around the highest saddle and at the exit point after quantum mechanical tunnelling of the fission barrier. Both choices correspond roughly to neutron-induced fission and spontaneous fission, respectively. We are allowed to treat such low energy system by the Langevin type dynamics as instead of the thermodynamical temperature in the Einstein relation Eq.~(\ref{Erel}), we take $T^*$ (Eq.~(\ref{Tstar})), which roughly describes the effect of the quantum mechanical fluctuations when $T\rightarrow 0$ \cite{WWH81, HKi98}. Of course, the initial configuration can not be sharp, so we have assumed that the beginning of each Langevin trajectory will be randomly distributed around the starting point, keeping the elongation $q^{\rm start}_2$=const and assuming that $q_3$ and $q_4$ and their conjugate momenta $p_3$ and $p_4$ will be randomly distributed around their starting value with the following condition:
\begin{equation}
\begin{array}{l}
 E_{\rm coll}=\displaystyle{V(q_3,q_4;q^{\rm start}_2)-V(q^{\rm start}_3,q^{\rm start}_3; q^{\rm start}_2)}\\[2ex]
~~~~~~~~+ \displaystyle{\frac{1}{2}\sum\limits_{i=3,4;j=3,4}{\cal M}_{ij}p_ip_j = E_0} ~,
\end{array}
\label{start}
\end{equation}
which assures the same initial collective energy ($E_{\rm coll}$) in each random trajectory. Here $E_0$ is the 'so-called' zero-point energy, equal to 1 MeV in our calculation. The system of Langevin equations has been solved using a discretization method in the time variable.

\begin{figure}[htb]
\centerline{\includegraphics[width=0.8\columnwidth]{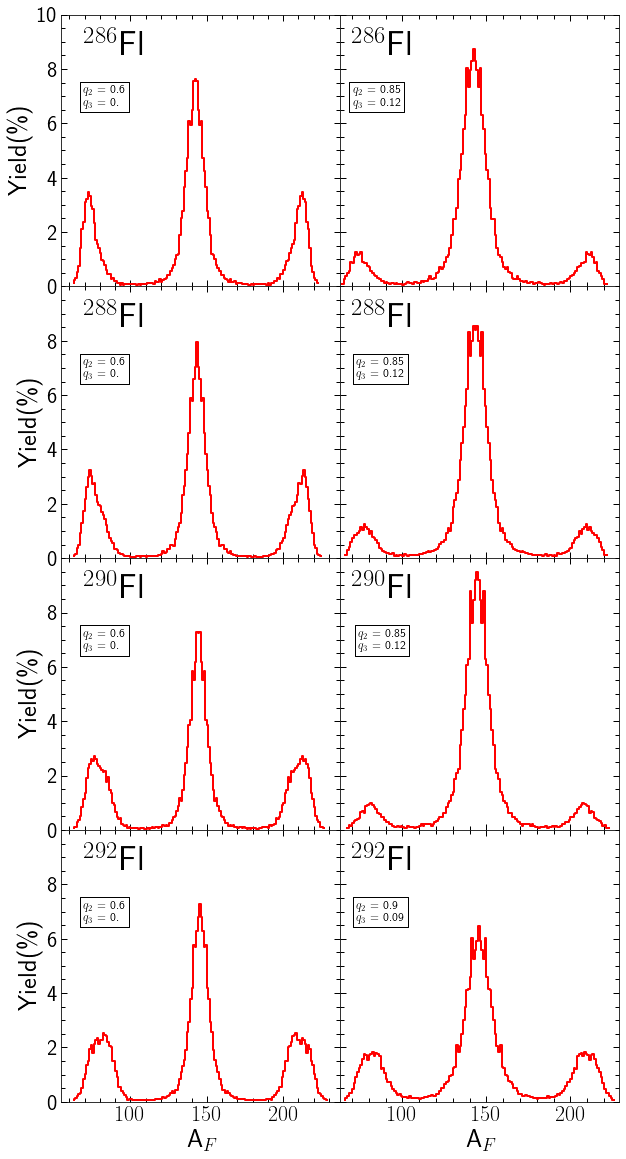}}
\caption{The same as in Fig.~\ref{Dsy} but for Fl isotopes.}
\label{Fly}
\end{figure}

The Langevin trajectory proceeds randomly towards fission within the following rectangular 3D box in the collective variables:
\begin{equation}
\begin{array}{l}
~~q^{start}_2\leq q_2 \\
-0.27\leq q_3 \leq 0.27\\
-0.21\leq q_4 \leq 0.21
\end{array}
\label{box}
\end{equation}
with reflecting walls that ensure that none of the trajectories will escape before reaching the scission configuration at larger elongations $q_2$. Of course, the box Eq.~(\ref{box}) is large enough, so the sticking of the walls occurs very rarely. A given trajectory ends when the neck radius of the fissioning nucleus reaches values smaller than 1 fm, which roughly corresponds to the 'size' of a nucleon. The time-step when solving the Langevin equations is taken $\Delta\tau=1\hbar/{\rm MeV}\approx\frac{2}{3}10^{-21}$ s. Typically 20.000 trajectories have to be generated to obtain sufficiently smooth the fission fragment mass yields as presented in Figs.~\ref{Dsy}-\ref{Fly}.

Our Langevin estimates of the fission fragment mass yields of $^{262}$Rf are compared in Fig.~\ref{Rf262} with the empirical data taken from Ref.~\cite{Lan96}. The agreement of the estimates with the data is satisfactory since none of the model parameters has been ''tuned'' to this data set in the region of SHN. Let us mention that their values are the same as admitted in the calculations in the actinides nuclei. It proves the choice of the Langevin calculation parameters is reasonable and gives a hope that the estimates for the heavier nuclei presented below are realistic too.

In Fig.~\ref{Dsy} the fission fragment mass yields of $^{270-282}$Ds isotopes are displayed. The l.h.s. columns show the yields corresponding to the case when the saddle point is taken as the starting point, while the r.h.s panel represented the mass yields when the exit point from the fission barrier was used as the starting point. In all presented figures, the symmetric fission picks dominate. However, two smaller picks corresponding to very asymmetric fission with the heavier fragment mass around A=208 are also visible. The asymmetric picks are significantly smaller when the exit from the barrier (i.e. spontaneous fission case) was taken as the starting point. This property of FMY is not astonishing since the higher excitation energy of the nuclear system offers a relatively larger probability of penetrating more exotic shape configurations exhibiting in peripheries of the PES.

Figs.~\ref{Cny} and \ref{Fly} show similar estimates of the FMY for $^{282-286}$Cn and $^{286-292}$Fl isotopes. The symmetric fission also dominates in these isotopes. Still, the contribution of the very asymmetric component becomes especially large (up to 40\% trajectories for Fl) when the Langevin trajectories begin around the saddle point. At the same time, in the spontaneous fission case, four times fewer trajectories lead to the asymmetric valley. Our estimates are in line with results obtained in Ref.~\cite{IZU20} where the 4D two-center shell model was used to evaluate the potential energy surfaces.

\section{Summary and conclusions}

Properties of superheavy elements with the charge number 104$\leq$ Z$\leq$126 were studied within the mac-mic model in the 4D Fourier deformation parameter space. All parameters of the present calculation are kept unchanged since they reproduce well the empirical masses and fission barrier heights of nuclei from different mass regions. 

The potential energy surfaces of 18 even-even isotope chains of elements from  Rf to Z=126 have been carefully studied, and the flooding technique was used to determine the fission barrier heights. An essential role of the non-axial and the left-right asymmetry degrees of freedom was shown when evaluating the barrier heights. The minimal energy path to fission of the heaviest spherical and oblate nuclei from this region frequently goes via very oblate and then non-axial shapes, leading to a significant decrease of their barrier heights.

The detailed investigation of the PES's suggests that the dominant fission channel of the SHN (apart from the lightest isotopes of Rf to Hs) is symmetric. In addition, in nuclei with A$\geq$280, another very asymmetric fission channel leading to the heavy fragment with mass A$\approx$208 appears. This effect seems to suggest smaller chances for the synthesis of elements with Z$\geq$118. 

At larger elongations ($q_2\geq 0.8$), the non-axial deformation may be neglected as the nucleus minimal energy configurations always become axially symmetric. That is why in our dissipative dynamics calculation, this deformation mode is not considered, and the Langevin equations were solved in the 3D space containing the elongation ($q_2$), neck ($q_4$), and mass asymmetry ($q_3$) degrees of freedom. Such a model describes well the main features of the fission fragment mass yields of superheavy nuclei. It was shown that adding higher-order terms ($q_5,\,q_6$) in the Fourier expansion Eqs.~(\ref{Fourier}, \ref{qi}) does not significantly change the PES properties. The Langevin calculation has allowed estimating the interplay between the symmetric and very asymmetric fission modes of the SHN. We have shown that in spontaneous fission, the fraction of trajectories leading to the heavy fragment with a mass around A=208 is a few times smaller than in the case of, e.g. neutron-induced fission where the Langevin trajectories begin in the vicinity of the saddle point. Both fission valleys, the symmetric and very asymmetric ones, are well separated at a larger elongation ($q_2$) of the nucleus. 

\vspace{0.5cm}

\noindent


%

\vfill

\section*{Appendix}

\begin{figure*}[h!]
\includegraphics[width=0.95\textwidth]{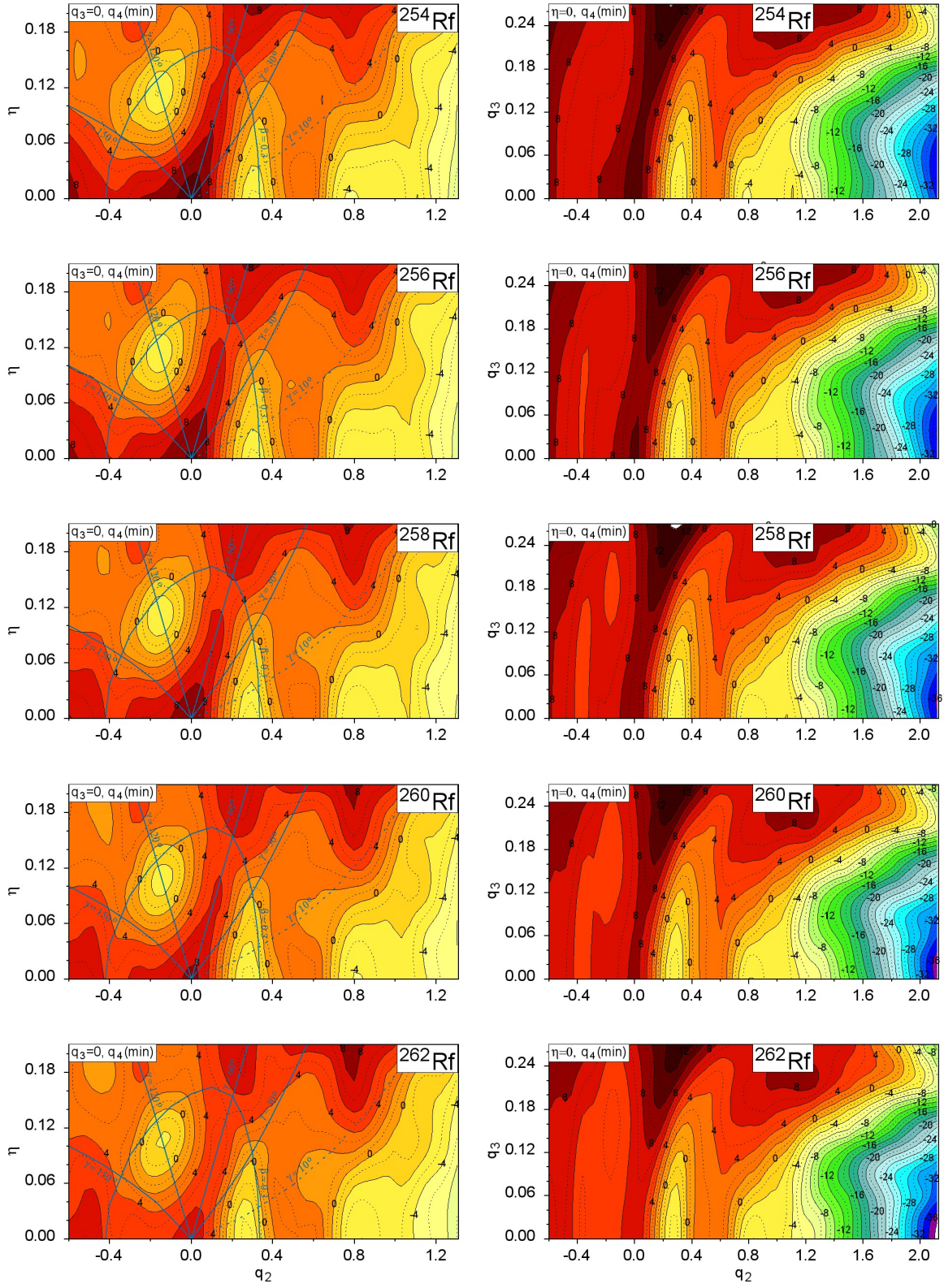}
\caption{Potential energy surface of $^{254-262}$Rf minimized with respect to  $q_4$ at the $(q_2,\eta)$ (l.h.s. column) and $(q_2,q_3)$ (r.h.s. column). The lines corresponding to $\gamma\,=\, 10^{\rm o},\,30^{\rm o},\,60^{\rm o},\,120^0$, and $150^{\rm o}$  as well as the line indicating to $\beta$=0.3 deformation are marked in the $(q_2,\eta)$  maps}
\label{Rf}
\end{figure*}

\begin{figure*}[h!]
\includegraphics[width=0.95\textwidth]{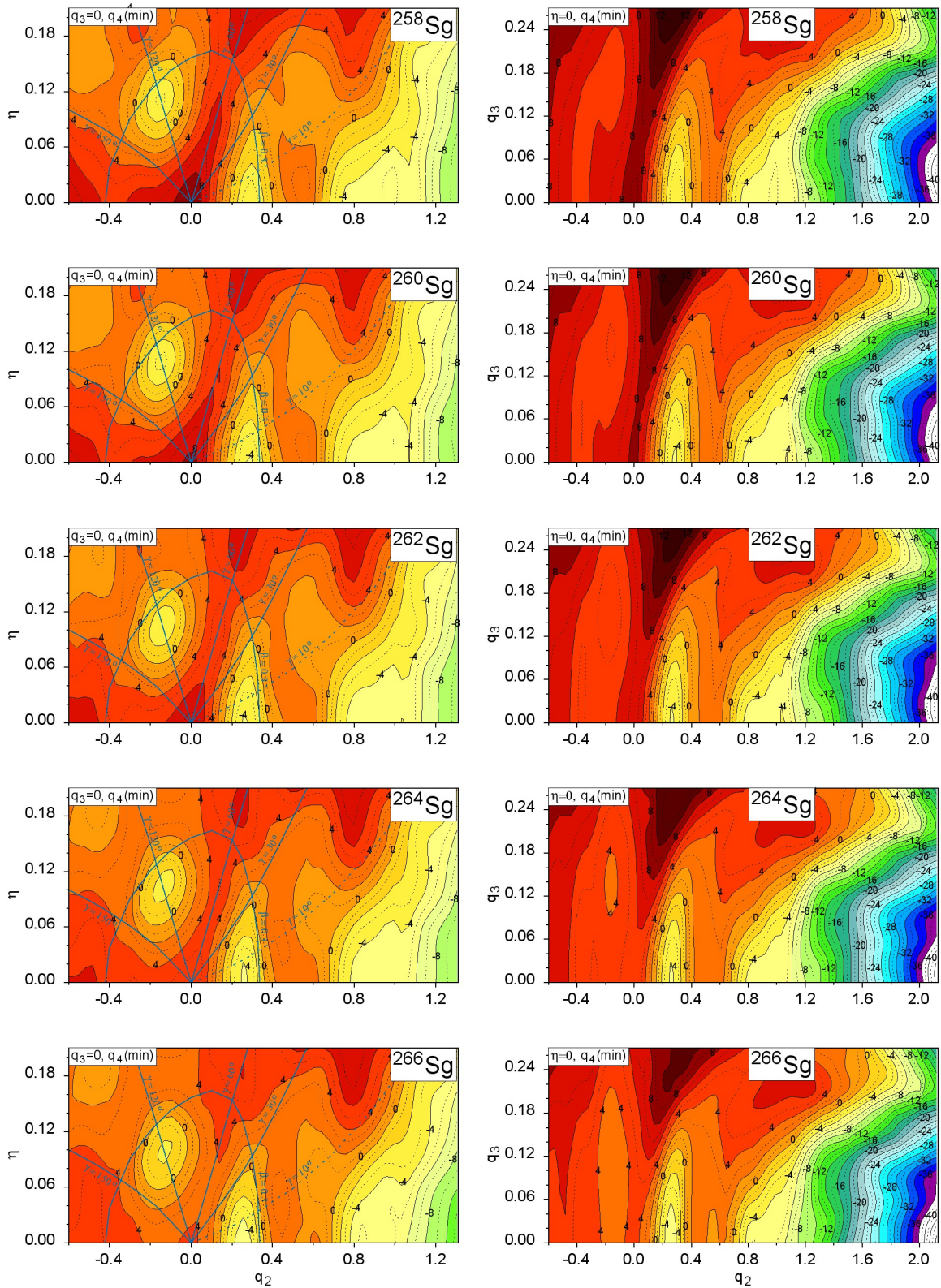}
\caption{The same as in Fig.~\ref{Rf} but for $^{258-266}$Sg isotopes.}
\label{Sg}
\end{figure*}

\begin{figure*}[h!]
\includegraphics[width=0.95\textwidth]{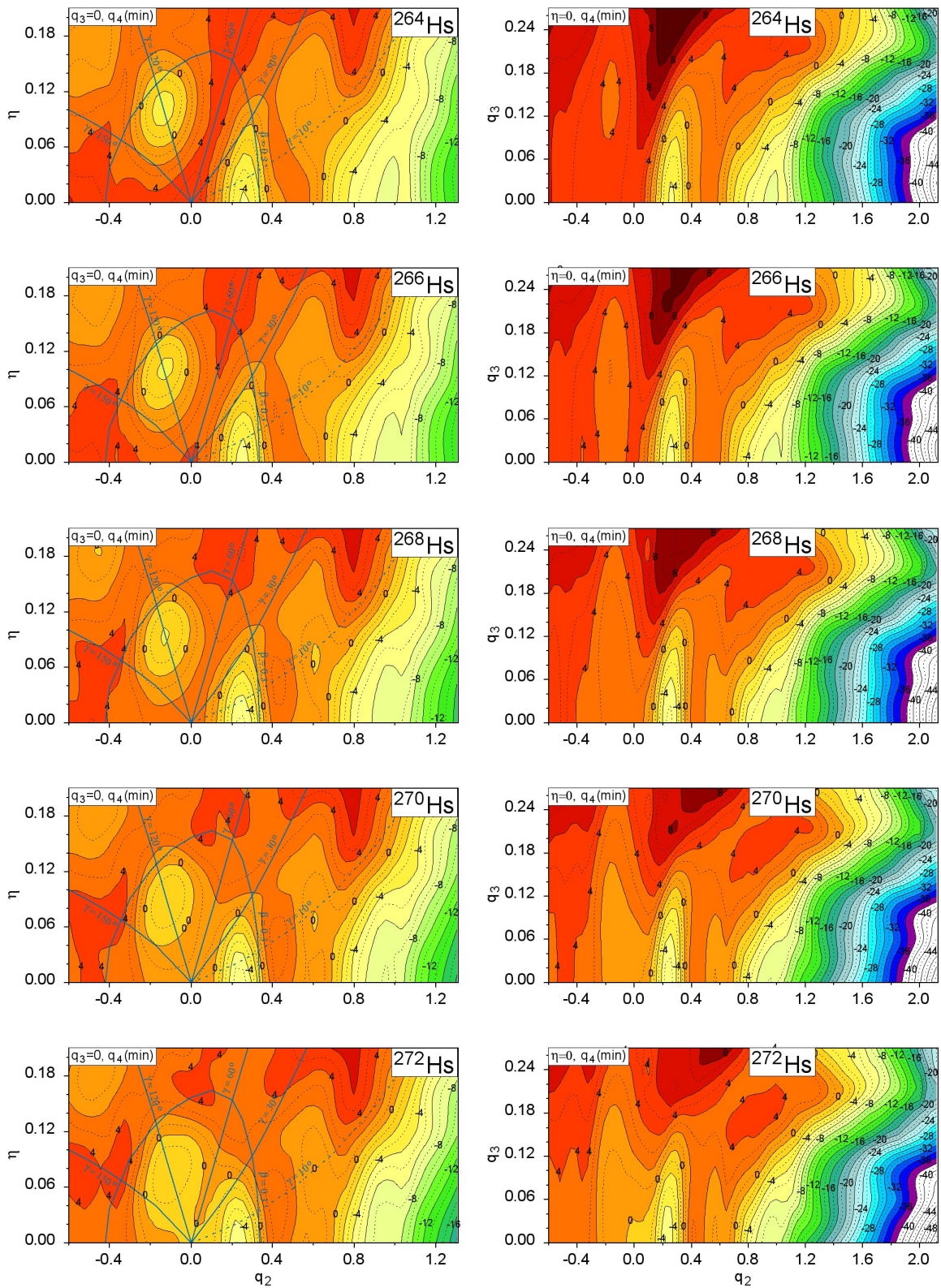}
\caption{The same as in Fig.~\ref{Rf} but for $^{264-272}$Hs isotopes.}
\label{Hs}
\end{figure*}

\begin{figure*}[h!]
\includegraphics[width=0.95\textwidth]{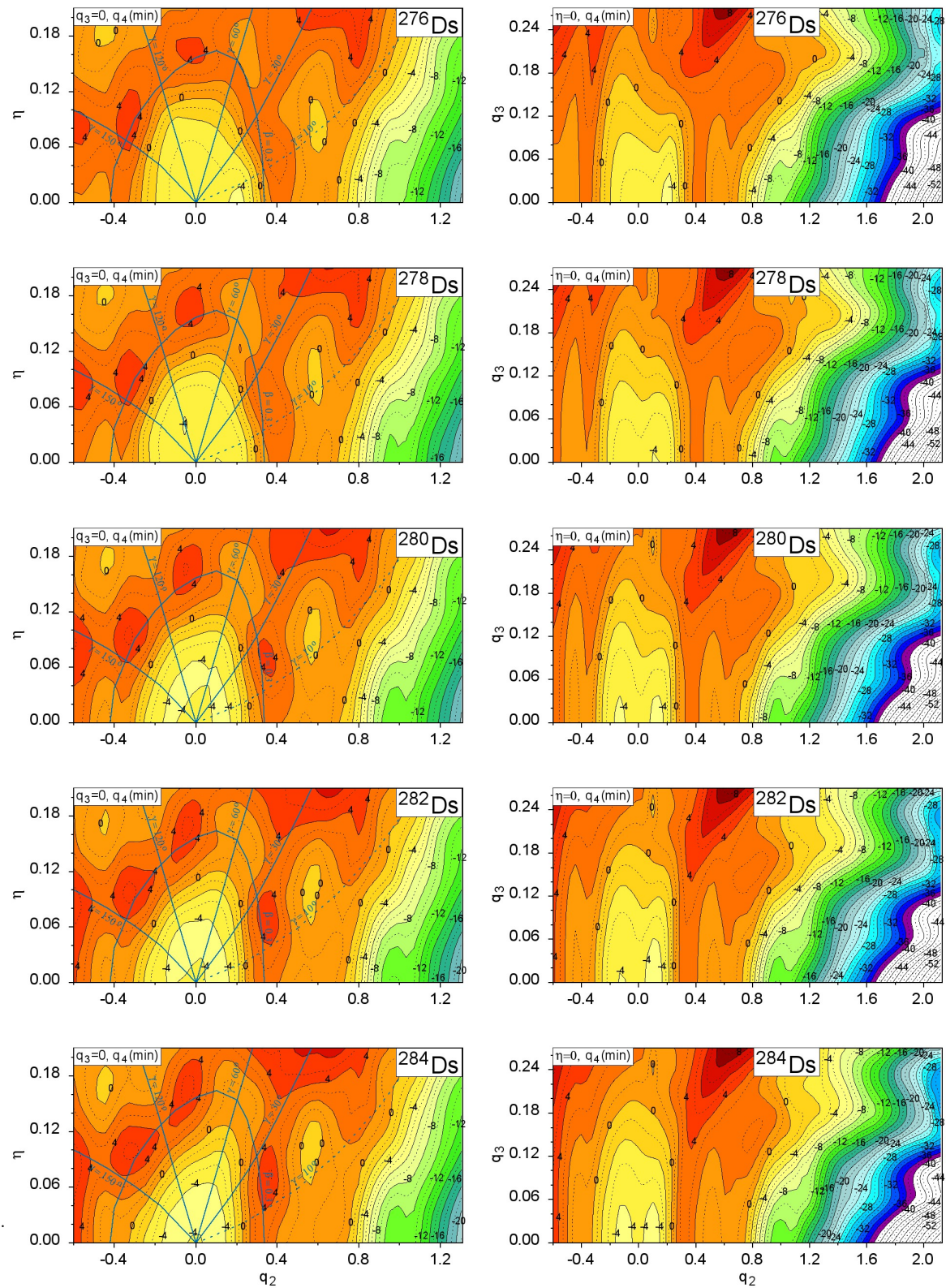}
\caption{The same as in Fig.~\ref{Rf} but for $^{276-284}$Ds isotopes.}
\label{Ds}
\end{figure*}

\begin{figure*}[h!]
\includegraphics[width=0.95\textwidth]{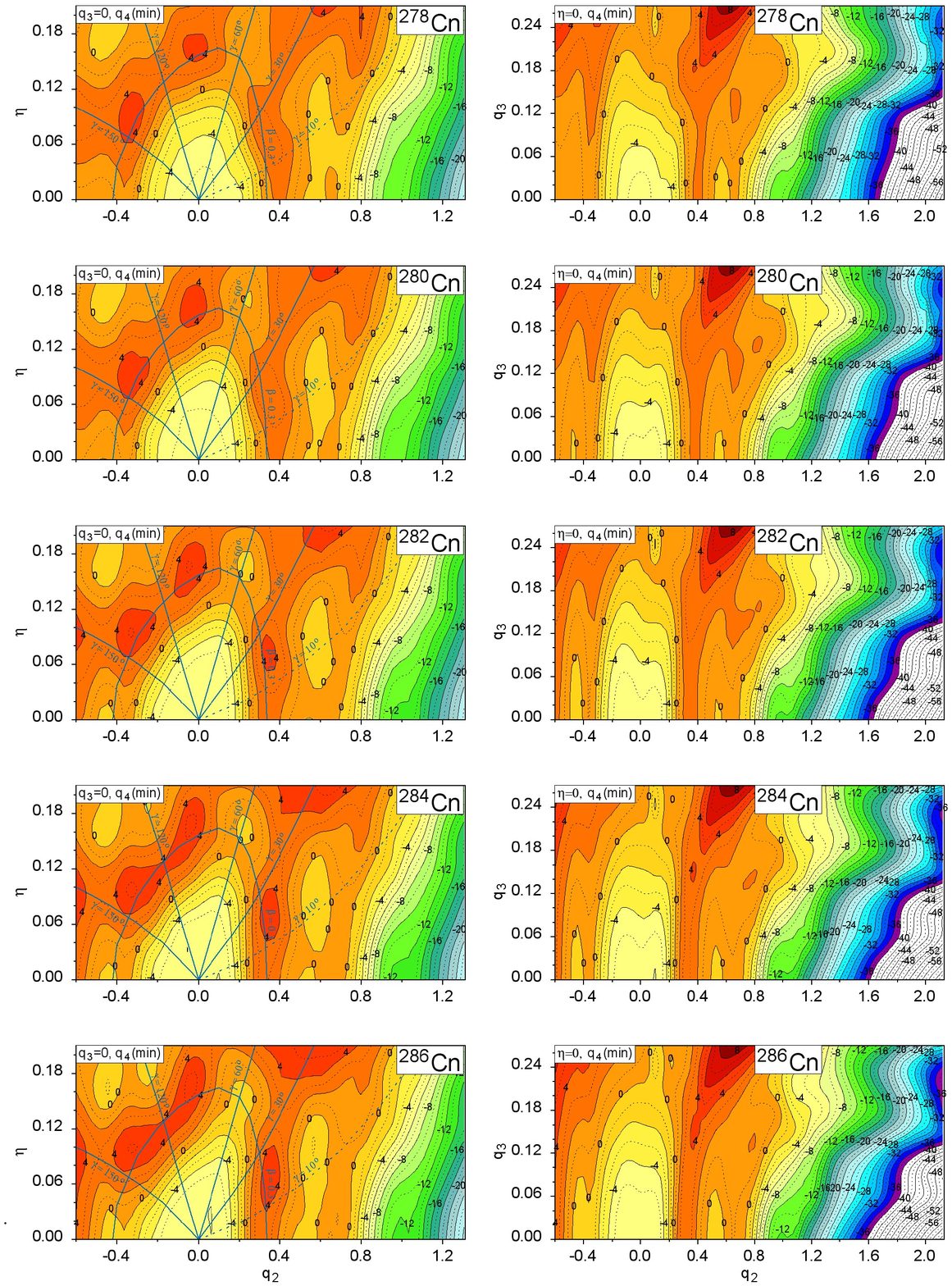}
\caption{The same as in Fig.~\ref{Rf} but for $^{}$Cn isotopes.}
\label{Cn}
\end{figure*}

\begin{figure*}[h!]
\includegraphics[width=0.95\textwidth]{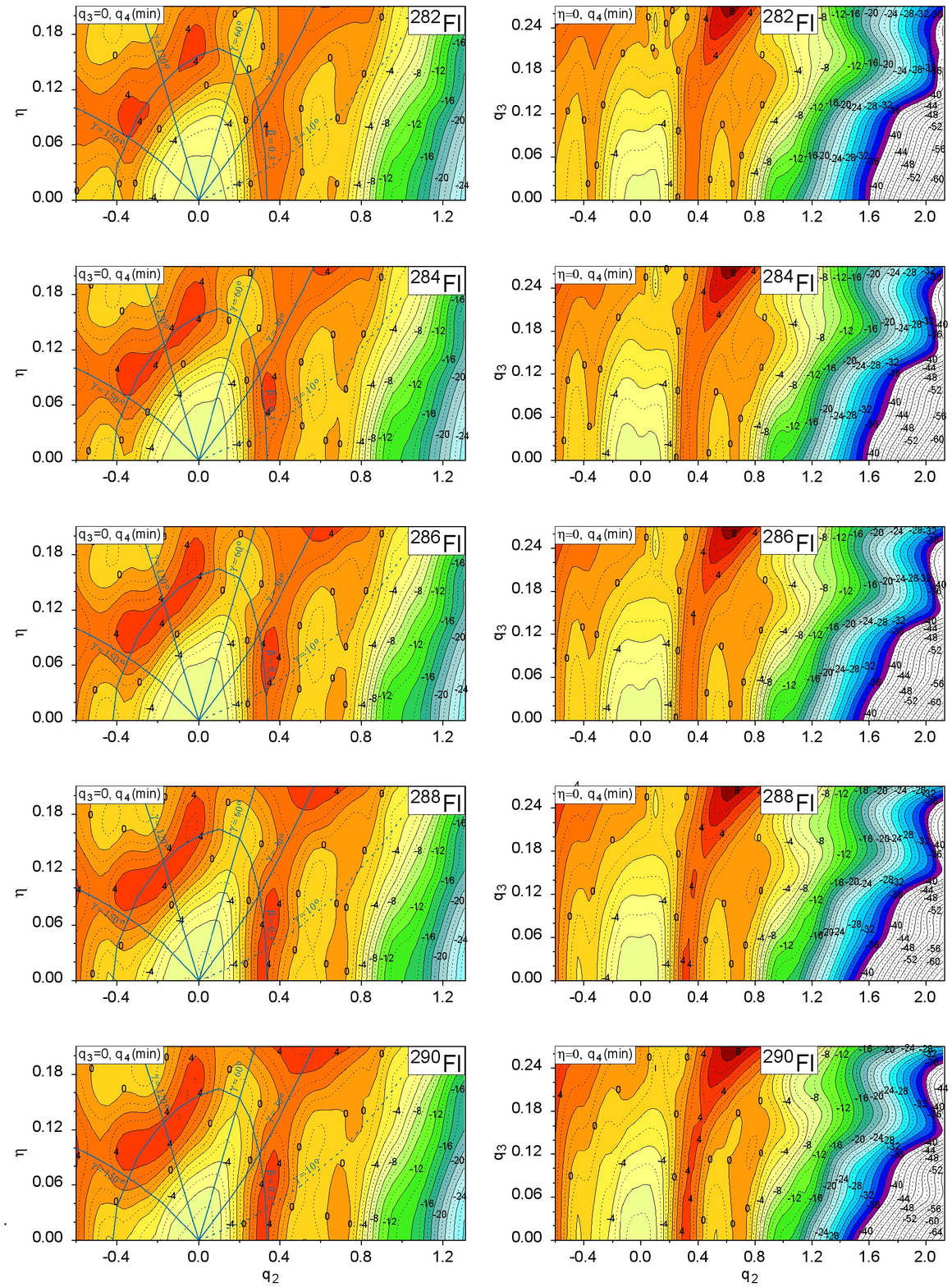}
\caption{The same as in Fig.~\ref{Rf} but for $^{282-290}$Fl isotopes.}
\label{Fl}
\end{figure*}

\begin{figure*}[h!]
\includegraphics[width=0.95\textwidth]{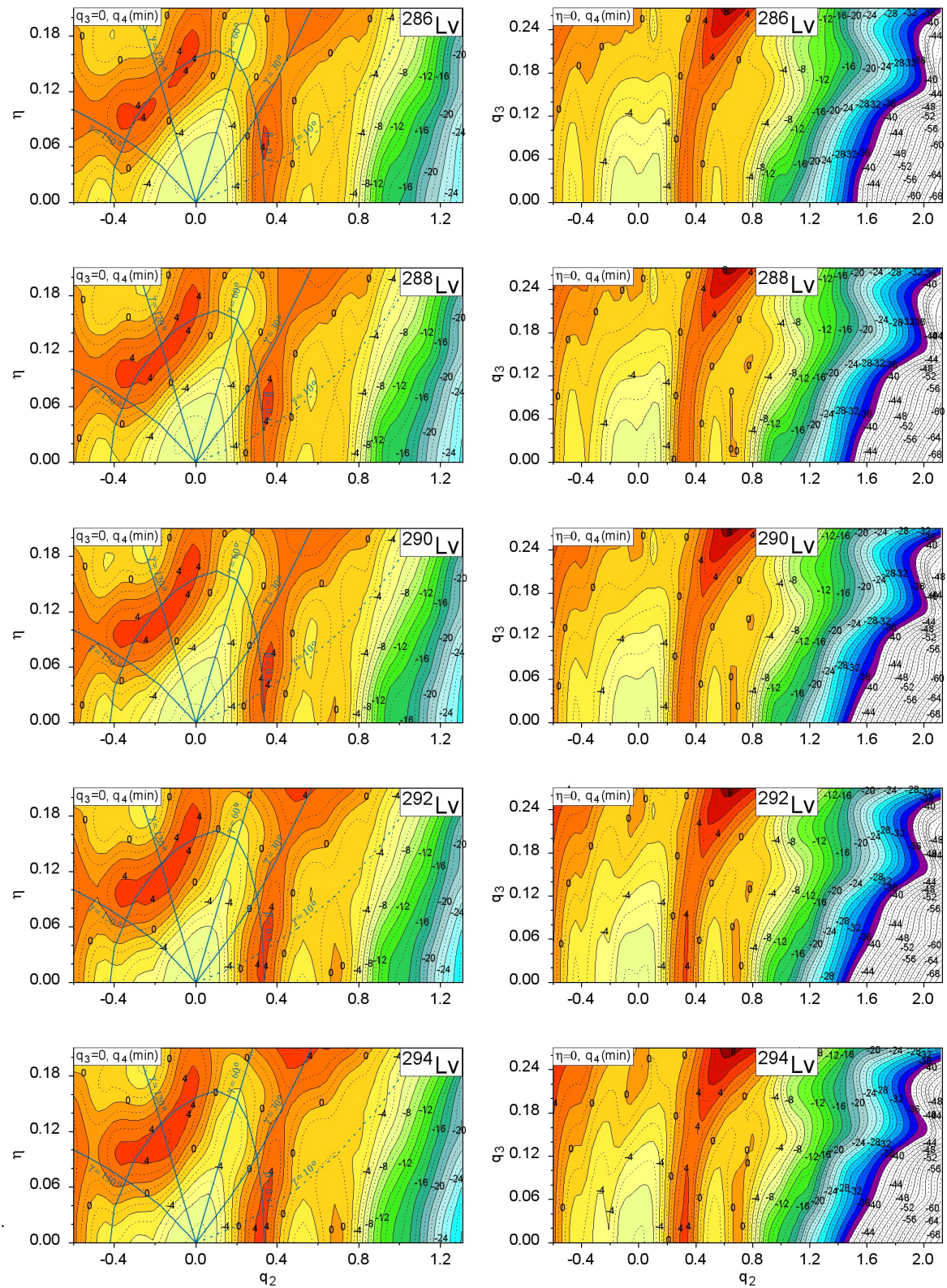}
\caption{The same as in Fig.~\ref{Rf} but for $^{286-294}$Lv isotopes.}
\label{Lv}
\end{figure*}

\begin{figure*}[h!]
\includegraphics[width=0.95\textwidth]{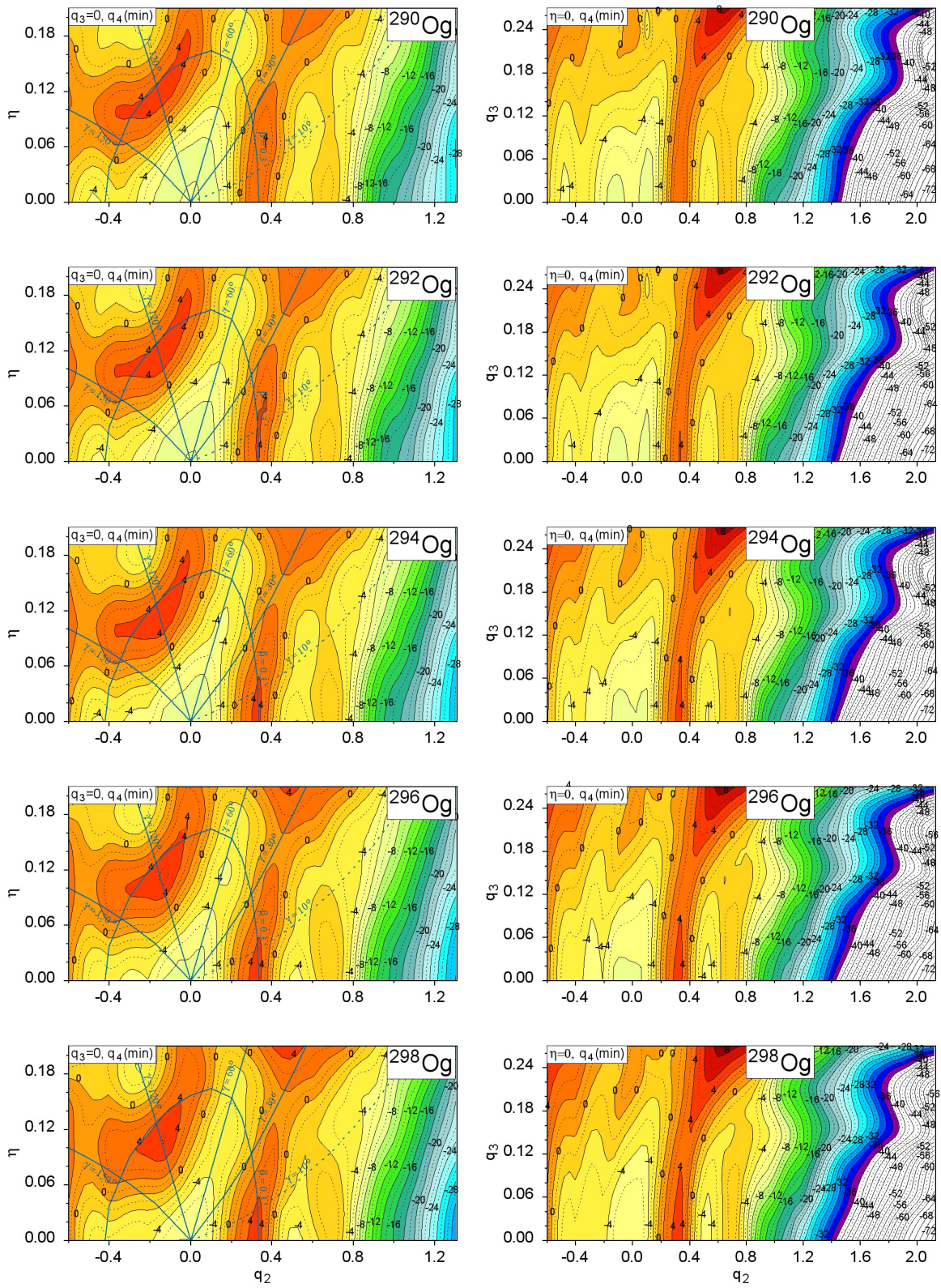}
\caption{The same as in Fig.~\ref{Rf} but for $^{290-298}$Og isotopes.}
\label{Og}
\end{figure*}

\begin{figure*}[h!]
\includegraphics[width=0.95\textwidth]{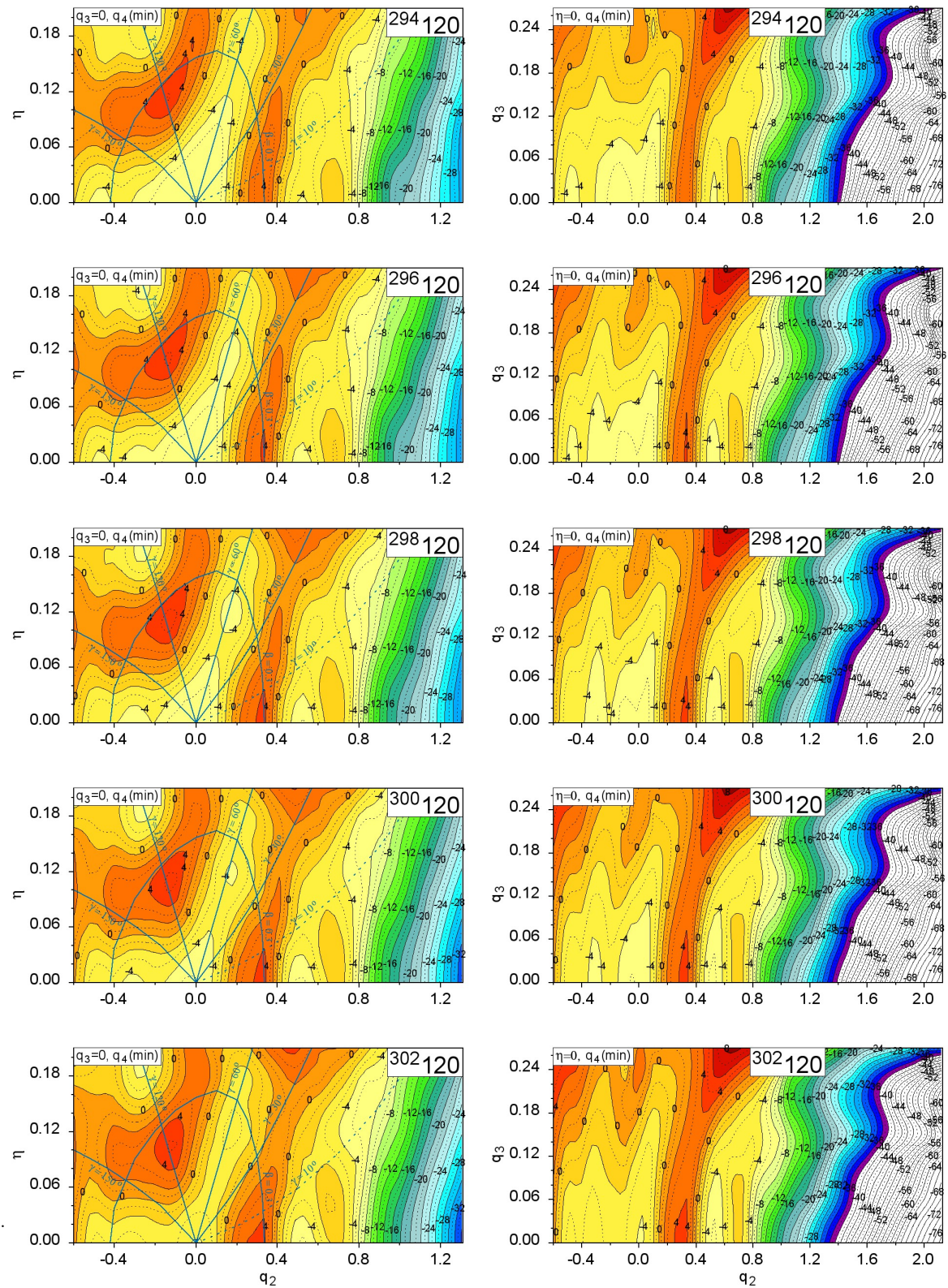}
\caption{The same as in Fig.~\ref{Rf} but for $^{294-302}$120 isotopes.}
\label{So}
\end{figure*}

In the following we present in Figs.~\ref{Rf}-\ref{So} the ($q_2,\eta$) (l.h.s. columns) and ($q_2,q_3$) (r.h.s. columns) cross-sections of the 4D potential energy surfaces of five selected even-even isotopes of elements with the charge number 104$\leq$Z$\leq$120. We believe that the plots can be useful for the experimentalists working in this region of nuclei. Of course, they are only 2D cross-sections, and they do not fully reflect the richness of the 4D PES. Nevertheless, one can see some main features like location of the ground state, shape isomers, possible shape coexistence, or different paths to fission at these plots. In Sec.~\ref{PES} we have described in detail the cross-sections of $^{304}$120 nucleus with the meaning of different curves envisaged in these maps, so we do not repeat here these explanations. 

As one can see in Figs.~\ref{Rf}-\ref{Hs} all presented isotopes of Rf, Sg, and Hs are prolate in the ground-state. It is due to the deformed shell effect reported already in 1990 by Patyk, and Sobiczewski \cite{PSo91}. The microscopic energy correction in the ground-state is around -3 MeV in Rf isotopes and becomes larger in Sg nuclei, reaching -4.5 MeV for $^{266}$Sg. The deformed shell energy effect in heavier Hs isotopes exceeds even -5 MeV. In Ds nuclei, one observes the decrease of the prolate deformation and appearance of the gamma-instability of the ground state in $^{280}$Ds and heavier Ds isotopes. At the same time, the microscopic energy correction exceeds -4 MeV in these nuclei. The $^{278-286}$Cn are typical transitional nuclei with a very flat PES around the spherical shape. In $^{286}$Cn isotope, the triaxial shape ($\gamma\approx 30^{\rm o}$) is preferred in the ground state. In Rf to Cn nuclei the axial($\gamma=0$) and non-axial ($\gamma\approx 30^{\rm o}$) paths to fission compete. The reduction of the saddle point energy due to non-axial deformation is less than 1 MeV in Rf isotopes and around 2 MeV in $^{258}$Sg. In $^{262-266}$Sg and all presented Hs isotopes, the axially symmetric and the non-axial fission barrier are comparable. In Ds and Cn isotopes a non-axial shape isomers appear at $q_2\approx 0.6$ and $\eta\approx 0.1$ (or $\gamma=10^{\rm o}$), while the axial and non-axial saddles have comparable energy. In Fl isotopes, the situation is similar. However, due to the left-right asymmetry ($q_3$) degree of freedom, the outer barrier of the non-axial isomers is diminished, so they have rather no chance to be populated. In $^{288-290}$Fl, Lv, Og, and Z=120 isotopes, the way to fission leads from the spherical, or nearly spherical, ground-state via oblate shapes and non-axial saddle to very elongated prolate deformations. A significant reduction (up to 3 MeV in $^{296-302}$120 nuclei) of the saddle point energy due to the $q_3$ deformation is observed in Lv and heavier elements. Apart the symmetric fission valley ($q_3\approx 0$) a very asymmetric valley ($q_3\approx 0.22$) appears in Ds and heavier elements.

\pagebreak
\pagebreak
\pagebreak
\pagebreak
\pagebreak
\pagebreak
\pagebreak
\pagebreak

\end{document}